\RequirePackage[2020-02-02]{latexrelease}
\documentclass[twocolumn,preprintnumbers]{revtex4}%
\usepackage{amssymb}
\usepackage{amsmath}
\usepackage{graphicx}
\usepackage{dcolumn}
\usepackage{bm}
\usepackage{xcolor}
\usepackage{ifthen}
\usepackage[normalem]{ulem}
\usepackage{amsfonts}%
\RequirePackage[2020-02-02]{latexrelease}
\UseRawInputEncoding
\providecommand{\U}[1]{\protect\rule{.1in}{.1in}}
\providecommand{\U}[1]{\protect\rule{.1in}{.1in}}
\def\showal{1}
\newcommand{\al}[1]{\ifthenelse{\showal=1}{\textcolor{orange}{[[#1]]}}{}}

\newcommand{\eb}[1]{\ifthenelse{\showal=1}{\textcolor{cyan}{[[#1]]}}{}}
\begin{document}
\title{Spontaneous collapse by entanglement suppression}
\author{Eyal Buks}
\email{eyal@ee.technion.ac.il}
\affiliation{Andrew and Erna Viterbi Department of Electrical Engineering, Technion, Haifa
32000, Israel}
\date{\today }

\begin{abstract}
We study a recently proposed modified Schr\"{o}dinger equation having an added
nonlinear term, which gives rise to disentanglement. The process of quantum
measurement is explored for the case of a pair of coupled spins. We find that
the deterministic time evolution generated by the modified Schr\"{o}dinger
equation mimics the process of wavefunction collapse. Added noise gives rise
to stochasticity in the measurement process. Conflict with both principles of
causality and separability can be avoided by postulating that the nonlinear
term is active only during the time when subsystems interact. Moreover, in the
absence of entanglement, all predictions of standard quantum mechanics are
unaffected by the added nonlinear term.

\end{abstract}
\maketitle





\textbf{Introduction} - In standard quantum mechanics a measurement is
described by a two-step process. The first step is governed by the standard
Schr\"{o}dinger equation. To avoid a possible paradoxical outcome of a
description based only on the first step (undefined cat state
\cite{Schrodinger_807}), a second step is postulated, in which the state
vector collapses. However, it has remained unknown how such a second step can
be self-consistently added \cite{Penrose_4864,Leggett_939,Leggett_022001}.
This difficulty has became known as the problem of quantum measurement.

In this work we explore an alternative to the collapse postulate, which is
based on a modified Schr\"{o}dinger equation that has an added nonlinear term
giving rise to disentanglement \cite{Buks_355303,Buks_025302}. The proposed
equation can be constructed for any physical system whose Hilbert space has
finite dimensionality, and it does not violate norm conservation  of the time
evolution. We explore the dynamics of a system made of two coupled spins, and
find that disentanglement gives rise to a process similar to state vector collapse.

Other types of nonlinear extensions of quantum mechanics
\cite{Geller_2200156} have been previously proposed and studied
\cite{Weinberg_336,Weinberg_61,Doebner_397,Doebner_3764,Gisin_5677,Kaplan_055002,Munoz_110503}%
. Most previously proposed extensions give rise to a spontaneous collapse
\cite{Bassi_471,Pearle_857,Ghirardi_470,Bassi_257,Arnquist_080401}. In some
cases, however, the proposed nonlinear models are inconsistent with
well-established physical principles. Moreover, many predictions of standard
quantum mechanics, that have been experimentally verified to very high
precision, are significantly altered by some of the proposed nonlinear
extensions. Such difficulties are discussed below in the final part of this
paper for the case of our proposed modified Schr\"{o}dinger equation. We find
that possible conflicts with the principles of causality and separability, and
with many experimentally confirmed predictions of standard quantum mechanics,
can be avoided by postulating that disentanglement is active only when
subsystems interact.

\begin{figure*}[ptb]
\begin{center}
\includegraphics[width=6in,keepaspectratio]{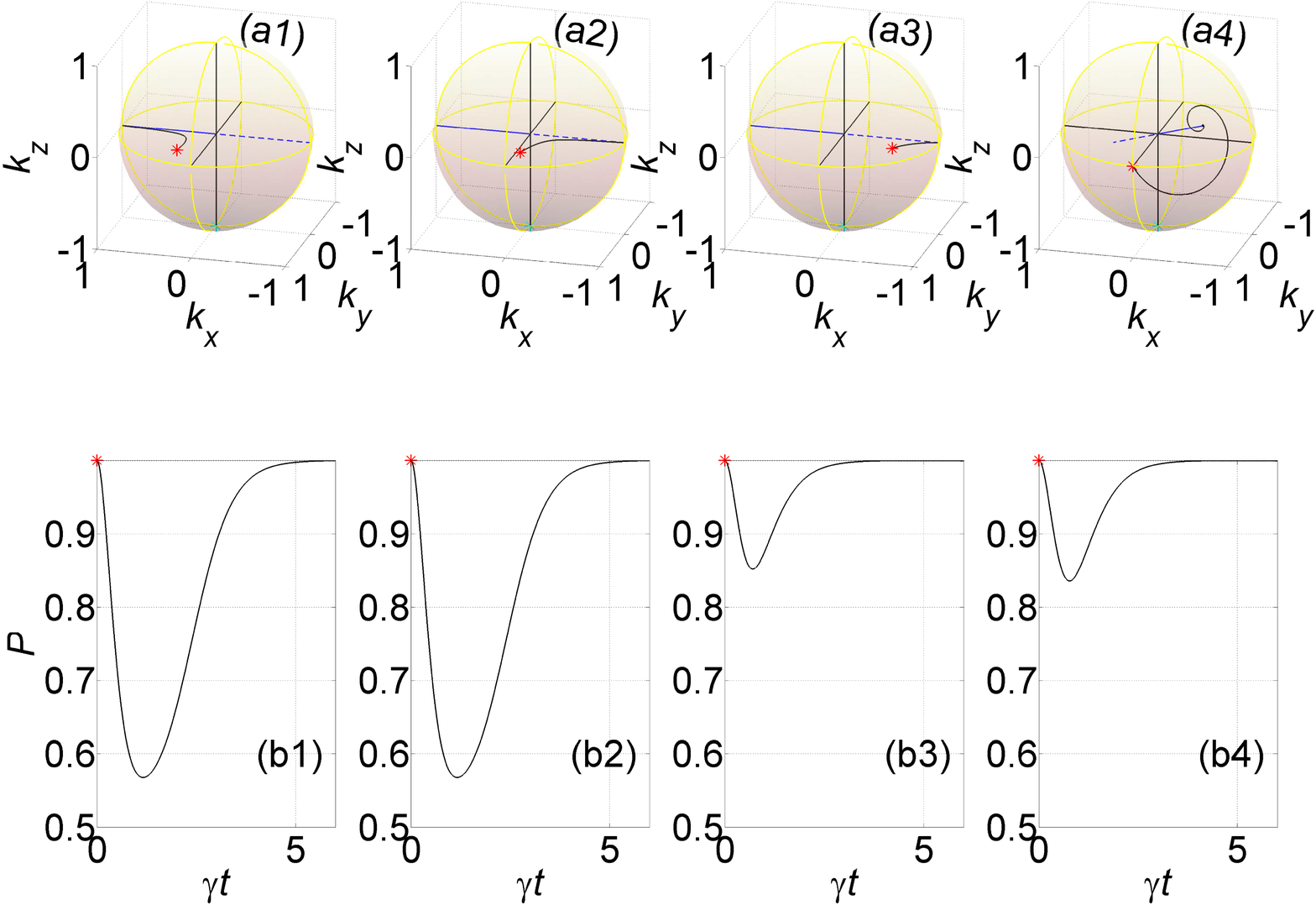}
\end{center}
\caption{Dipolar measurement. The spin numbers are $S_{1}=1/2$ and
$S_{2}=21/2$, the rates are $\gamma=\omega_{\mathrm{d}}=1$, and $\mathbf{\hat
{n}}_{2}=-\mathbf{\hat{z}}$ (initial direction of the $S_{2}$ spin, which is
labeled by a cyan star symbols). For the plots labeled by the numbers 1, 2 and
3 the dipolar unit vector $\mathbf{\hat{u}}_{\mathrm{d}}$ is given by
$\mathbf{\hat{u}}_{\mathrm{d}}=\mathbf{\hat{x}}$ (i.e. $\mathbf{\hat{u}%
}_{\mathrm{d}}$ is perpendicular to $\mathbf{\hat{n}}_{2}$), whereas
$\mathbf{\hat{u}}_{\mathrm{d}}=\left(  \sin\left(  3\pi/8\right)  \cos\left(
3\pi/4\right)  ,\sin\left(  3\pi/8\right)  \sin\left(  3\pi/4\right)
,\cos\left(  3\pi/8\right)  \right)  $ for the plots labeled by the number 4
(i.e. $\mathbf{\hat{n}}_{2}\cdot\mathbf{\hat{u}}_{\mathrm{d}}\neq0$). At time
$t=0$ the spin 1/2 is pointing in the direction $\mathbf{\hat{n}}_{1}=\left(
\sin\theta_{1}\cos\varphi_{1},\sin\theta_{1}\sin\varphi_{1},\cos\theta
_{1}\right)  $, where for (1) $\theta_{1}=0.55\pi$ and $\varphi_{1}=0.45\pi$,
for (2) $\theta_{1}=0.55\pi$ and $\varphi_{1}=0.55\pi$, for (3) $\theta
_{1}=0.55\pi$ and $\varphi_{1}=0.75\pi$, and for (4) $\theta_{1}=0.5\pi$, and
$\varphi_{1}=0.5\pi$. Red star symbols label the initial points $\mathbf{\hat
{n}}_{1}$, and the blue solid (dashed) lines connect the origin and the unit
vectors $\mathbf{\hat{u}}_{\mathrm{d}}$ ($-\mathbf{\hat{u}}_{\mathrm{d}}$).}%
\label{Fig2SpinDE}%
\end{figure*}

\begin{figure}[ptb]
\begin{center}
\includegraphics[width=3in,keepaspectratio]{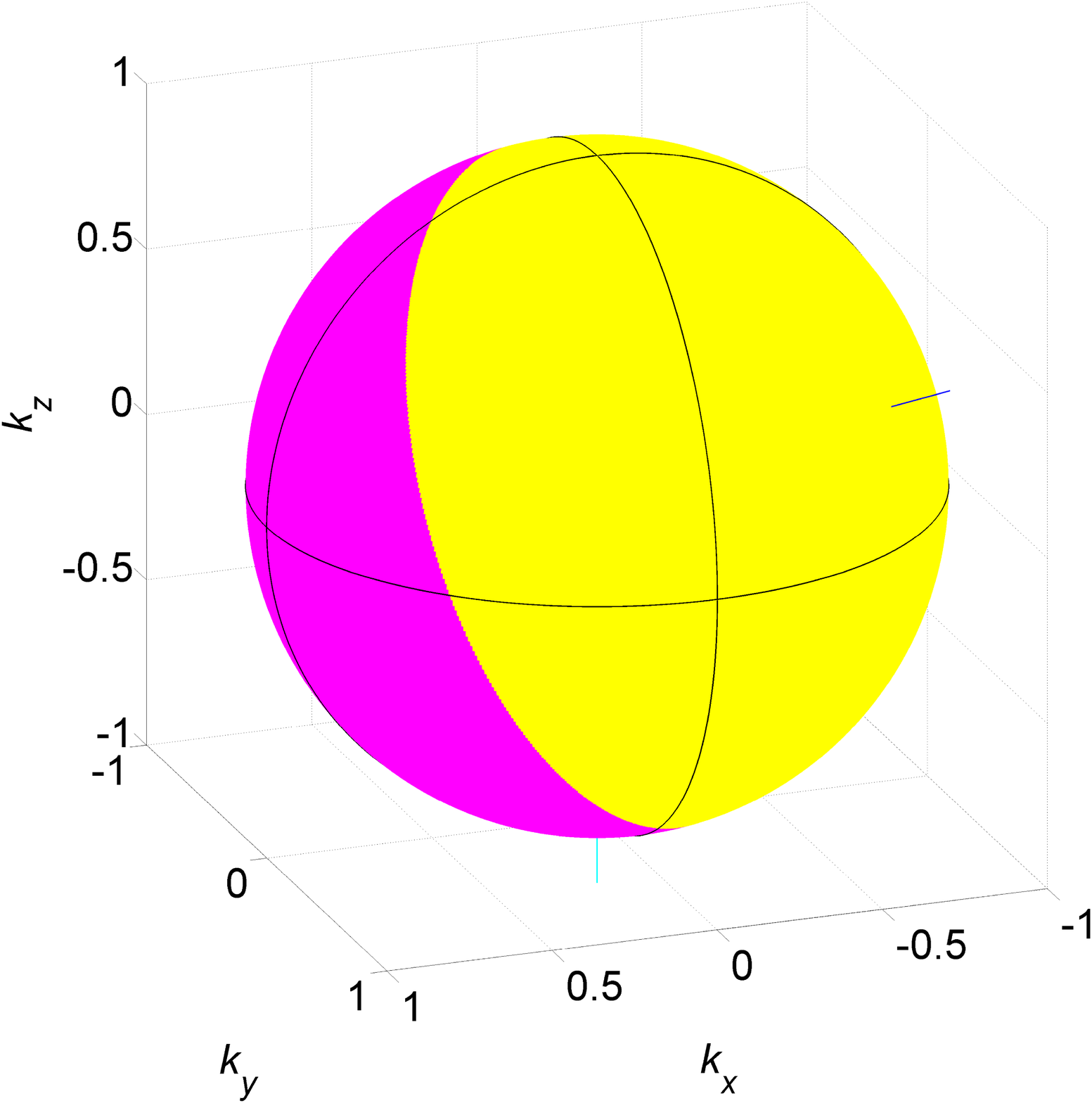}
\end{center}
\caption{Basins of attraction. All parameters are the same as those used to
generate the plots of Fig. \ref{Fig2SpinDE} labeled by the number 4. Initial
direction of the $S_{2}$ spin $\mathbf{\hat{n}}_{2}=-\mathbf{\hat{z}}$ is
labeled by a cyan tick, and the dipolar unit vector $\mathbf{\hat{u}%
}_{\mathrm{d}}=\left(  \sin\left(  3\pi/8\right)  \cos\left(  3\pi/4\right)
,\sin\left(  3\pi/8\right)  \sin\left(  3\pi/4\right)  ,\cos\left(
3\pi/8\right)  \right)  $ is labeled by a blue tick. At time $t=0$ the spin
1/2 is pointing in the direction $\mathbf{\hat{n}}_{1}$. The yellow (purple)
colored region is the basin of attraction lying in the hemisphere
$\mathbf{\hat{n}}_{1}\cdot\mathbf{\hat{u}}_{\mathrm{d}}>0$ ($\mathbf{\hat{n}%
}_{1}\cdot\mathbf{\hat{u}}_{\mathrm{d}}<0$), and the corresponding attractor
is $\mathbf{\hat{u}}_{\mathrm{d}}$ ($-\mathbf{\hat{u}}_{\mathrm{d}}$).}%
\label{FigBOA}%
\end{figure}

\textbf{Disentanglement} - Consider a system composed of two subsystems
labeled as '1' and '2', respectively. The dimensionality of the Hilbert spaces
of both subsystems, which is denoted by $N_{1}$ and $N_{2}$, respectively, is
assumed to be finite. The system is in a normalized pure state vector
$\left\vert \psi\right\rangle $ given by%
\begin{equation}
\left\vert \psi\right\rangle =\mathcal{K}_{1}C\otimes\mathcal{K}%
_{2}^{\mathrm{T}}\;, \label{psi C}%
\end{equation}
where $C$ is a $N_{1}\times N_{2}$ matrix having entries $C_{k_{1},k_{2}}$,
matrix transposition is denoted by $\mathrm{T}$, $\mathcal{K}_{1}=\left(
\left\vert k_{1}\right\rangle _{1},\left\vert k_{2}\right\rangle _{1}%
,\cdots,\left\vert k_{N_{1}}\right\rangle _{1}\right)  $, $\mathcal{K}%
_{2}=\left(  \left\vert k_{1}\right\rangle _{2},\left\vert k_{2}\right\rangle
_{2},\cdots,\left\vert k_{N_{2}}\right\rangle _{2}\right)  $, and $\left\{
\left\vert k_{1}\right\rangle _{1}\right\}  $ ($\left\{  \left\vert
k_{2}\right\rangle _{2}\right\}  $) is an orthonormal basis spanning the
Hilbert space of subsystem '1' ('2').

The purity $P_{1}$ ($P_{2}$) is defined by $P_{1}=\operatorname{Tr}\rho
_{1}^{2}$ ($P_{2}=\operatorname{Tr}\rho_{2}^{2}$), where $\rho_{1}%
=\operatorname{Tr}_{2}\rho$\ ($\rho_{2}=\operatorname{Tr}_{1}\rho$) is the
reduced density operator of the first (second) subsystem. By employing the
Schmidt decomposition one finds that $P_{1}=P_{2}\equiv P$, where
$P=1-\left\langle \mathcal{Q}\right\rangle =1-\left\langle \psi\right\vert
\mathcal{Q}\left\vert \psi\right\rangle $, the operator $\mathcal{Q}$ is given
by [see Eq. (\ref{1-P V2}) of appendix A, and Ref. \cite{Buks_QMLN}]%
\begin{equation}
\mathcal{Q}=\frac{1}{2}%
{\displaystyle\sum\limits_{k_{1}^{\prime}<k_{1}^{\prime\prime}}}
{\displaystyle\sum\limits_{k_{2}^{\prime}<k_{2}^{\prime\prime}}}
\left\vert \Psi_{k_{1}^{\prime},k_{1}^{\prime\prime},k_{2}^{\prime}%
,k_{2}^{\prime\prime}}\right\rangle \left\langle \Psi_{k_{1}^{\prime}%
,k_{1}^{\prime\prime},k_{2}^{\prime},k_{2}^{\prime\prime}}\right\vert \;,
\label{Q operator}%
\end{equation}
and the state $\left\langle \Psi_{k_{1}^{\prime},k_{1}^{\prime\prime}%
,k_{2}^{\prime},k_{2}^{\prime\prime}}\right\vert $, which depends on the
matrix $C$ corresponding to a given state $\left\vert \psi\right\rangle $, is
given by (note that $\left\langle \Psi_{k_{1}^{\prime},k_{1}^{\prime\prime
},k_{2}^{\prime},k_{2}^{\prime\prime}}\right\vert $ is not normalized)%
\begin{equation}
\left\langle \Psi_{k_{1}^{\prime},k_{1}^{\prime\prime},k_{2}^{\prime}%
,k_{2}^{\prime\prime}}\right\vert =C_{_{\mathrm{d}}}\left\langle
\mathrm{a}\right\vert +C_{_{\mathrm{a}}}\left\langle \mathrm{d}\right\vert
-C_{\mathrm{c}}\left\langle \mathrm{b}\right\vert -C_{\mathrm{b}}\left\langle
\mathrm{c}\right\vert \;, \label{Psi bra}%
\end{equation}
where $\mathrm{a}=k_{1}^{\prime},k_{2}^{\prime}$, $\mathrm{b}=k_{1}^{\prime
},k_{2}^{\prime\prime}$, $\mathrm{c}=k_{1}^{\prime\prime},k_{2}^{\prime}$ and
$\mathrm{d}=k_{1}^{\prime\prime},k_{2}^{\prime\prime}$. Note that
$\left\langle \mathcal{Q}\right\rangle =0$ for a product state. In standard
quantum mechanics $\left\langle \mathcal{Q}\right\rangle $ is time independent
when the subsystems are decoupled (i.e. their mutual interaction vanishes).

As an example, consider a two spin 1/2 system (i.e. $N_{1}=N_{2}=2$) in a pure
state $\left\vert \psi\right\rangle $ given by $\left\vert \psi\right\rangle
=a\left\vert --\right\rangle +b\left\vert -+\right\rangle +c\left\vert
+-\right\rangle +d\left\vert ++\right\rangle $. For this case the sum in Eq.
(\ref{Q operator}) contains a single term with $\left\langle \Psi\right\vert
=d\left\langle -,-\right\vert -c\left\langle -,+\right\vert -b\left\langle
+,-\right\vert +a\left\langle +,+\right\vert $, and thus $P=1-2\left\vert
ad-bc\right\vert ^{2}$. Note that for this case $\left\langle \mathcal{Q}%
\right\rangle \leq1/2$ (provided that $\left\vert \psi\right\rangle $ is
normalized) \cite{Wootters_2245}.

Consider a modified Schr\"{o}dinger equation for the ket vector $\left\vert
\psi\right\rangle $ having the form%
\begin{equation}
\frac{\mathrm{d}}{\mathrm{d}t}\left\vert \psi\right\rangle =\left[
-i\hbar^{-1}\mathcal{H}-\gamma\left(  \mathcal{Q}-\left\langle \mathcal{Q}%
\right\rangle \right)  \right]  \left\vert \psi\right\rangle \;, \label{NLSE}%
\end{equation}
where $\hbar$ is the Planck's constant, $\mathcal{H}^{{}}=\mathcal{H}^{\dag}$
is the Hamiltonian, the rate $\gamma$ is positive, and the operator
$\mathcal{Q}$ is given by Eq. (\ref{Q operator}). The added nonlinear term
proportional to $\gamma$ gives rise to disentanglement, however, it has no
effect when $\left\vert \psi\right\rangle $ represents a product state. Note
that the norm conservation condition $0=\left(  \mathrm{d}/\mathrm{d}t\right)
\left\langle \psi\right.  \left\vert \psi\right\rangle $ is satisfied by the
modified Schr\"{o}dinger equation (\ref{NLSE}).

\textbf{Dipolar interaction} - As an example, the dynamics generated by the
modified Schr\"{o}dinger equation (\ref{NLSE}) is explored for the case of
dipolar interaction between two spins having spin quantum numbers $S_{1}$ and
$S_{2}$, respectively. The dipolar interaction is represented by the operator
$V_{\mathrm{d}}=\hbar^{-1}\omega_{\mathrm{d}}\left(  \mathbf{S}_{1}%
\cdot\mathbf{\hat{u}}_{\mathrm{d}}\right)  \left(  \mathbf{S}_{2}%
\cdot\mathbf{\hat{u}}_{\mathrm{d}}\right)  $, where the rate $\omega
_{\mathrm{d}}$ is positive, $\mathbf{S}_{n}$ is the spin angular momentum
vector operator of the $n$'th spin ($n\in\left\{  1,2\right\}  $), and
$\mathbf{\hat{u}}_{\mathrm{d}}=\left(  \sin\theta\cos\varphi,\sin\theta
\sin\varphi,\cos\theta\right)  $ is a unit vector.

Time evolution examples for the case $S_{1}=1/2$ and $S_{2}=21/2$ are shown by
the plots in Fig. \ref{Fig2SpinDE}. The initial state at time $t=0$ is a
product state, for which the spin 1/2 is pointing in the direction of the unit
vector $\mathbf{\hat{n}}_{1}$ (labeled by a red star symbol), and the spin
21/2 is pointing in the direction of the unit vector $\mathbf{\hat{n}}%
_{2}=-\mathbf{\hat{z}}$ (labeled by a cyan star symbol). The overlaid blue
solid (dashed) lines connect the origin and the dipolar coupling unit vectors
$\mathbf{\hat{u}}_{\mathrm{d}}$ ($-\mathbf{\hat{u}}_{\mathrm{d}}$). The spin
1/2 Bloch vector $\mathbf{k}=\left(  \hbar/2\right)  ^{-1}\left\langle
\mathbf{S}_{1}\right\rangle $ is numerically calculated by integrating the
modified Schr\"{o}dinger equation (\ref{NLSE}) for the case $\mathcal{H}%
=V_{\mathrm{d}}$. The black solid lines in Fig. \ref{Fig2SpinDE}(a1), (a2),
(a3) and (a4) represent the spin 1/2 Bloch vector $\mathbf{k}$ evolving from
its initial value $\mathbf{\hat{n}}_{1}$ at time $t=0$. The single-spin purity
$P=1-\left\langle \mathcal{Q}\right\rangle $ as a function of time $t$ is
shown in Fig. \ref{Fig2SpinDE}(b1), (b2), (b3) and (b4).

For the plots in Fig. \ref{Fig2SpinDE} labeled by the numbers 1, 2 and 3, the
dipolar unit vector $\mathbf{\hat{u}}_{\mathrm{d}}$ is given by $\mathbf{\hat
{u}}_{\mathrm{d}}=\mathbf{\hat{x}}$ (i.e. $\mathbf{\hat{u}}_{\mathrm{d}}$ is
perpendicular to $\mathbf{\hat{n}}_{2}=-\mathbf{\hat{z}}$). These plots, which
differ by the initial direction $\mathbf{\hat{n}}_{1}$ of the spin 1/2
(labeled by red star symbols), demonstrate that the Bloch sphere is divided
into two basins of attraction. The first (second) basin is the hemisphere
$\mathbf{\hat{n}}_{1}\cdot\mathbf{\hat{u}}_{\mathrm{d}}>0$ ($\mathbf{\hat{n}%
}_{1}\cdot\mathbf{\hat{u}}_{\mathrm{d}}<0$), and the corresponding attractor
is $\mathbf{\hat{u}}_{\mathrm{d}}$ ($-\mathbf{\hat{u}}_{\mathrm{d}}$).

\begin{figure}[ptb]
\begin{center}
\includegraphics[width=3in,keepaspectratio]{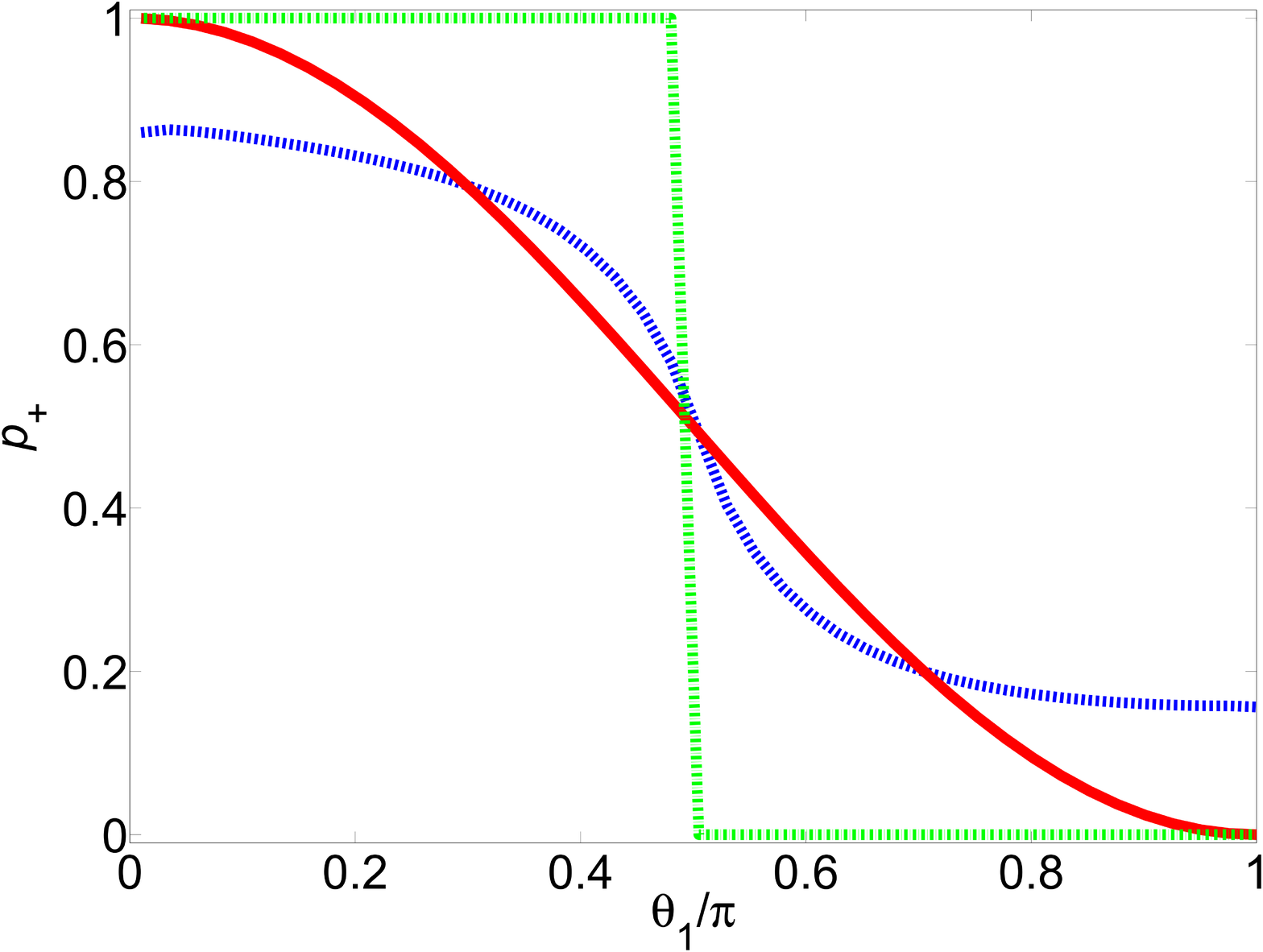}
\end{center}
\caption{Noise. The probability $p_{+}$ is plotted as a function of the angle
$\theta_{1}=\cos^{-1}\left(  \mathbf{\hat{n}}_{1}\cdot\mathbf{\hat{u}
}_{\mathrm{d}}\right)  $ for the noiseless case (green dash-dotted line), the
case $\phi_{0}=0.5$ (blue dashed line), and the Born rule (red solid line).}%
\label{FigN}%
\end{figure}

While $\mathbf{\hat{n}}_{2}\cdot\mathbf{\hat{u}}_{\mathrm{d}}=0$ for the plots
in Fig. \ref{Fig2SpinDE} labeled by the numbers 1, 2 and 3, the behavior when
the initial spin $S_{2}$ direction $\mathbf{\hat{n}}_{2}$ is not perpendicular
to the dipolar coupling unit vector $\mathbf{\hat{u}}_{\mathrm{d}}$ is
demonstrated by the plots labeled by the number 4. The plot in Fig.
\ref{Fig2SpinDE}(a4) shows that the Bloch vector trajectory, from the initial
value $\mathbf{\hat{n}}_{1}$ (labeled by the red star symbol) towards the
attractor at $\mathbf{\hat{u}}_{\mathrm{d}}$ becomes spiral-like when
$\mathbf{\hat{n}}_{2}\cdot\mathbf{\hat{u}}_{\mathrm{d}}\neq0$. The basins of
attraction for this case (i.e. plots in Fig. \ref{Fig2SpinDE} labeled by the
number 4) are shown in Fig. \ref{FigBOA}. This example demonstrates that the
dipolar unit vector $\mathbf{\hat{u}}_{\mathrm{d}}$ determines the spin 1/2
component that is being measured. The measurement process is deterministic
however the outcome, which is either $+1$ (when $\mathbf{\hat{n}}_{1}%
\cdot\mathbf{\hat{u}}_{\mathrm{d}}>0$) or $-1$ (when $\mathbf{\hat{n}}%
_{1}\cdot\mathbf{\hat{u}}_{\mathrm{d}}<0$) is quantized. This behavior is
demonstrated by the green dash-dotted line in Fig. \ref{FigN}, in which the
probability $p_{+}$ that the measurement outcome is $+1$ is plotted as a
function of the angle $\theta_{1}=\cos^{-1}\left(  \mathbf{\hat{n}}_{1}%
\cdot\mathbf{\hat{u}}_{\mathrm{d}}\right)  $. For comparison, the red solid
line represents the Born rule of standard quantum mechanics, for which
$p_{+}\left(  \theta_{1}\right)  =\cos^{2}\left(  \theta_{1}/2\right)  $. A
simplified model is employed below to explore noise-induced stochasticity.

\textbf{Noise} - The effect of external noise is taken into account by
applying a random rotation to the initial spin 1/2 Block vector $\mathbf{\hat
{n}}_{1}$. The random rotation is characterized by an axis normal to
$\mathbf{\hat{n}}_{1}$, and by a rotation angle $\phi_{\mathrm{r}}$.  As an
example, consider the case where the rotation angle $\phi_{\mathrm{r}}$ has a
wrapped Cauchy probability distribution $f\left(  \phi_{\mathrm{r}}\right)  $
given by%
\begin{equation}
f\left(  \phi_{\mathrm{r}}\right)  =\frac{1}{2\pi}\frac{\sinh\phi_{0}}%
{\cosh\phi_{0}-\cos\phi_{\mathrm{r}}}\;,
\end{equation}
where $\phi_{0}>0$ is a scale factor. Consider a rotated frame, in which the
dipolar unit vector $\mathbf{\hat{u}}_{\mathrm{d}}$ is parallel to the unit
vector $\mathbf{\hat{z}}$. The unit vector $\mathbf{\hat{n}}_{1}^{{}}$ in this
frame is denoted by $\mathbf{\hat{n}}_{1\mathrm{R}}$. The probability $p_{+}$
that the measurement outcome is $+1$ is calculated by spherical integration
over the hemisphere $z^{\prime}\geq0$%
\begin{equation}
p_{+}=\frac{1}{4\pi}\int_{0}^{\pi/2}\mathrm{d}\theta^{\prime}\;\sin
\theta^{\prime}\int_{0}^{2\pi}\mathrm{d}\varphi^{\prime}\;\frac{4f\left(
\theta_{1\mathrm{R}}\right)  }{\sin\theta_{1\mathrm{R}}}\;, \label{p_+}%
\end{equation}
where $\theta_{1\mathrm{R}}=\cos^{-1}\left(  \mathbf{\hat{n}}_{1\mathrm{R}%
}\cdot\mathbf{\hat{n}}^{\prime}\right)  $, and where $\mathbf{\hat{n}}%
^{\prime}=\left(  \sin\theta^{\prime}\cos\varphi^{\prime},\sin\theta^{\prime
}\sin\varphi^{\prime},\cos\theta^{\prime}\right)  $. As can be seen from the
blue dashed line in Fig. \ref{FigN}, which is calculated using Eq. (\ref{p_+})
with a scale factor of $\phi_{0}=0.5$, noise-induced stochasticity mimics the
behavior predicted by the Born rule (red solid line).

\textbf{The measurement time} - For the examples shown in Fig.
\ref{Fig2SpinDE}, initially at time $t=0$, the ket vector $\left\vert
\psi\right\rangle $ represents a product state having single-spin purity
$P=1$. The time dependency of $P$ is shown in Fig. \ref{Fig2SpinDE}(b1), (b2),
(b3) and (b4). In the short time limit of $\omega_{\mathrm{d}}t\ll1$ the
effect of the disentanglement term in the modified Schr\"{o}dinger equation
(\ref{NLSE}) is relatively weak (since $\left\langle \mathcal{Q}\right\rangle
$ is initially small), and consequently $P$ rapidly drops due to entanglement
generated by the dipolar interaction $V_{\mathrm{d}}$. At latter times, when
disentanglement becomes sufficiently efficient, the single-spin purity $P$
starts increasing. Interaction-induced generation of entanglement becomes
inefficient when the spin 1/2 becomes nearly parallel or nearly anti-parallel
to the dipolar unit vector $\mathbf{\hat{u}}_{\mathrm{d}}$, and consequently
the single-spin purity $P$ approaches unity in the long time limit.

For sufficiently short times after turning on the interaction (i.e. after
$t=0$), time evolution is dominated by the effect of the dipolar interaction.
When the effect of the disentanglement term is disregarded, one finds that in
the short time limit the following holds $\mathrm{d}\left\langle
\mathbf{S}_{n}\right\rangle /\mathrm{d}t\simeq\omega_{n}\mathbf{\hat{u}%
}_{\mathrm{d}}\times\left\langle \mathbf{S}_{n}\right\rangle $, where
$n\in\left\{  1,2\right\}  $, $\omega_{1}=\omega_{\mathrm{d}}\hbar
^{-1}\left\langle \mathbf{S}_{2}\cdot\mathbf{\hat{u}}_{\mathrm{d}%
}\right\rangle $ and $\omega_{2}=\omega_{\mathrm{d}}\hbar^{-1}\left\langle
\mathbf{S}_{1}\cdot\mathbf{\hat{u}}_{\mathrm{d}}\right\rangle $. Thus, in the
short time limit, the purity $P$ is roughly given by $P\simeq1-\left(
2^{-3/2}S_{2}\left\vert \mathbf{\hat{n}}_{1}\times\mathbf{\hat{u}}%
_{\mathrm{d}}\right\vert \left(  \mathbf{\hat{n}}_{2}\cdot\mathbf{\hat{u}%
}_{\mathrm{d}}\right)  \omega_{\mathrm{d}}t\right)  ^{2}$ [see Eqs. (6.192)
and (8.701) of Ref. \cite{Buks_QMLN}, and note that it is assumed that in the
short time limit the spin states are nearly spin coherent states
\cite{Radcliffe_313}]. The above-derived expression for the purity time
evolution $P\left(  t\right)  $ reveals the dependence of short-time dynamics
on the macroscopicity of the measuring apparatus (i.e. the second spin), which
is represented by the spin number $S_{2}$.

\textbf{Vanishing Hamiltonian} - To gain further insight into the
disentanglement process generated by the nonlinear term $-\gamma\left(
\mathcal{Q}-\left\langle \mathcal{Q}\right\rangle \right)  $ added to the
Schr\"{o}dinger equation (\ref{NLSE}), consider for simplicity the case where
the Hamiltonian vanishes, i.e. $\mathcal{H}=0$. The Schmidt decomposition of a
general state vector $\left\vert \psi\right\rangle $ is expressed as
\begin{equation}
\left\vert \psi\right\rangle =%
{\displaystyle\sum\limits_{l=1}^{\min\left(  N_{1},N_{2}\right)  }}
q_{l}\left\vert l,l\right\rangle \;, \label{Schmidt decomposition}%
\end{equation}
where $q_{l}$ are non-negative real numbers, the tensor product$\ \left\vert
l\right\rangle _{1}\otimes\left\vert l\right\rangle _{2}$ is denoted by
$\left\vert l,l\right\rangle \ $, and $\left\{  \left\vert l\right\rangle
_{1}\right\}  $ ($\left\{  \left\vert l\right\rangle _{2}\right\}  $) is an
orthonormal basis spanning the Hilbert space of subsystem '1' ('2'). Note that
for a product state $q_{l}=\delta_{l,l_{0}}$, where $l_{0}\in\left\{
1,2,\cdots,\min\left(  N_{1},N_{2}\right)  \right\}  $. The normalization
condition reads $\left\langle \psi\right.  \left\vert \psi\right\rangle
=L_{2}=1$, where the $n$'th moment $L_{n}$ is defined by%
\begin{equation}
L_{n}=%
{\displaystyle\sum\limits_{l=1}^{\min\left(  N_{1},N_{2}\right)  }}
q_{l}^{n}\;. \label{L_n}%
\end{equation}
Note that for a product state $L_{n}=1$ for any positive integer $n$ (provided
that $\left\vert \psi\right\rangle $ is normalized).

\begin{figure}[ptb]
\begin{center}
\includegraphics[width=3in,keepaspectratio]{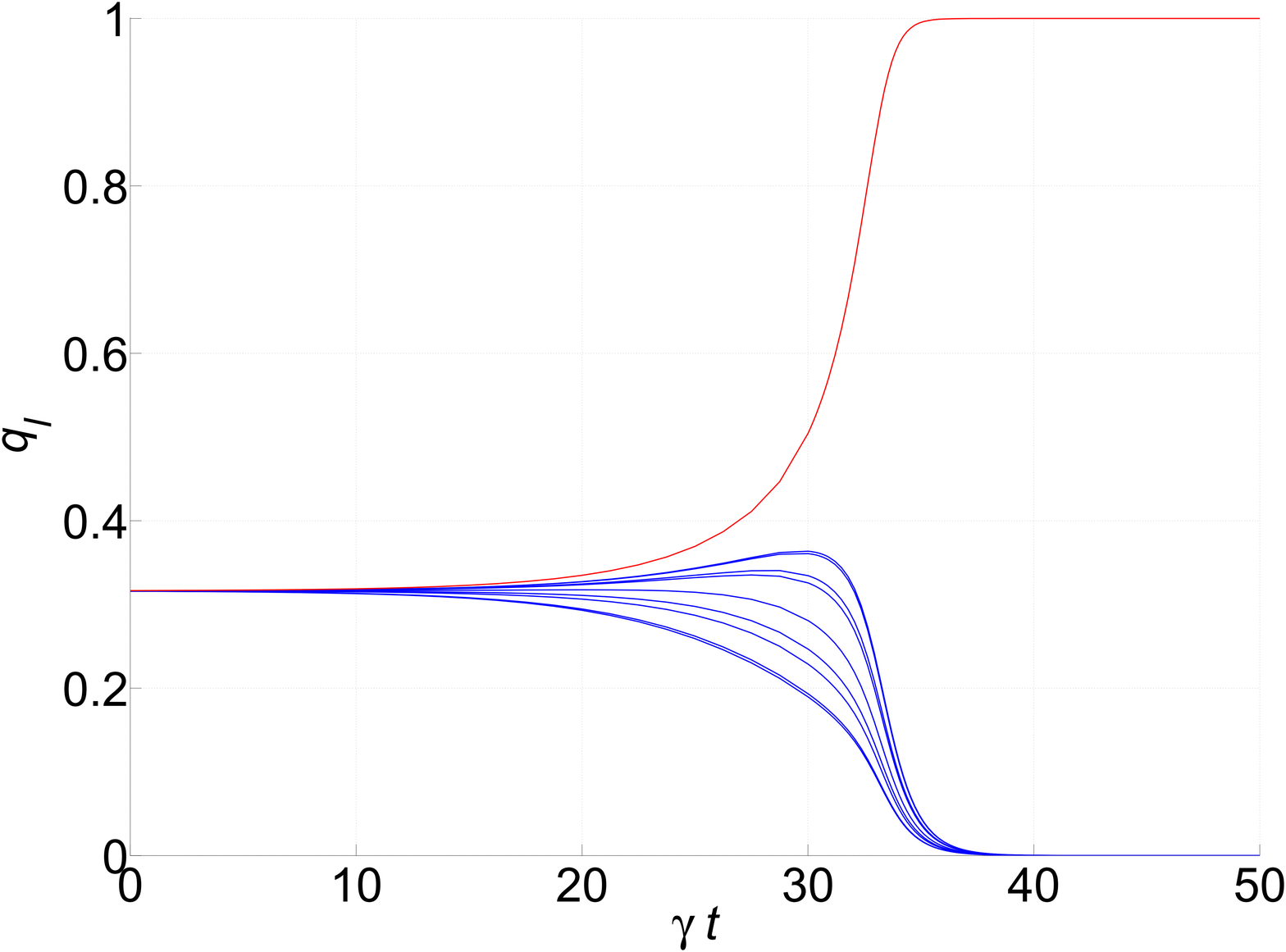}
\end{center}
\caption{Vanishing Hamiltonian. The plot shows an example solution of the set
of equations (\ref{d log q_l / dt}) for the case $\min\left(  N_{1}%
,N_{2}\right)  =10$ and $\gamma=1$. The solution for $q_{l_{0}}\left(
t\right)  $ is represented by the red line, whereas the blue lines represent
the solutions for $q_{l}\left(  t\right)  $ with $l\neq l_{0}$. For this
example, $q_{l}\left(  t=0\right)  \simeq\left(  \min\left(  N_{1}%
,N_{2}\right)  \right)  ^{-1/2}$, i.e. the initial value of the purity $P$ is
close to its smallest possible value of $1/\min\left(  N_{1},N_{2}\right)  $.
The corresponding initial entropy $\sigma$ is close to its largest possible
value of $\log\left(  \min\left(  N_{1},N_{2}\right)  \right)  $. In the limit
$t\rightarrow\infty$ the purity $P\rightarrow1$ (largest possible value) and
the entropy $\sigma\rightarrow0$ (smallest possible value).}%
\label{Figdqdt}%
\end{figure}

In the Schmidt basis, the following holds [see Eqs. (\ref{Q operator}) and
(\ref{Psi bra})]%
\begin{equation}
\mathcal{Q}\left\vert \psi\right\rangle =%
{\displaystyle\sum\limits_{l=1}^{\min\left(  N_{1},N_{2}\right)  }}
q_{l}\left(  1-q_{l}^{2}\right)  \left\vert l,l\right\rangle \;,
\end{equation}
and $\left\langle \mathcal{Q}\right\rangle =1-L_{4}$, and thus [see Eq.
(\ref{NLSE})]%
\begin{equation}
\frac{\mathrm{d\log}q_{l}}{\mathrm{d}t}=\gamma\left(  q_{l}^{2}-L_{4}\right)
\;. \label{d log q_l / dt}%
\end{equation}
An example solution of the set of equations (\ref{d log q_l / dt}) for the
case $\min\left(  N_{1},N_{2}\right)  =10$ and $\gamma=1$ is shown in Fig.
\ref{Figdqdt}.

The time evolution of the $n$'th moment $L_{n}$ is governed by [see Eqs.
(\ref{L_n}) and (\ref{d log q_l / dt})]%
\begin{equation}
\frac{\mathrm{d}L_{n}}{\mathrm{d}t}=n\gamma\left(  L_{n+2}-L_{n}L_{4}\right)
\;. \label{d L_n / dt}%
\end{equation}
For the case of $n=2$, Eq. (\ref{d L_n / dt}) yields the norm conservation
condition $0=\left(  \mathrm{d}/\mathrm{d}t\right)  \left\langle \psi\right.
\left\vert \psi\right\rangle $, which is satisfied provided that $\left\vert
\psi\right\rangle $ is normalized, i.e. $L_{2}=1$ [see Eq.
(\ref{Schmidt decomposition})]. For the case $n=4$ Eq. (\ref{d L_n / dt})
yields an evolution equation for the purity $P=L_{4}$, which is given by
$\mathrm{d}L_{4}/\mathrm{d}t=4\gamma\left(  L_{6}-L_{4}^{2}\right)  $. Using
the Cauchy--Schwarz inequality one finds that $L_{4}^{2}\leq L_{2}L_{6}$ [see
Eq. (\ref{L_n})], hence $\mathrm{d}P/\mathrm{d}t\geq0$ (recall the
normalization condition $L_{2}=1$), i.e. the purity $P$ monotonically
increases with time. The same conclusion can alternatively be  drawn from Eq.
(\ref{d log q_l / dt}), which can be expressed as $\mathrm{d}q_{l}%
/\mathrm{d}t=\partial H/\partial q_{l}$, where $H=\left(  \gamma/4\right)
\left(  3-2L_{2}\right)  L_{4}$\ [see Eq. (\ref{L_n}), and note that
$H=\left(  \gamma/4\right)  L_{4}=\left(  \gamma/4\right)  P$ when $L_{2}=1$].

For any two integers $l^{\prime},l^{\prime\prime}\in\left\{  1,2,\cdots
,\min\left(  N_{1},N_{2}\right)  \right\}  $ the following holds [see Eq.
(\ref{d log q_l / dt})]%
\begin{equation}
\frac{\mathrm{d\log}\frac{q_{l^{\prime}}}{q_{l^{\prime\prime}}}}{\mathrm{d}%
t}=\gamma\left(  q_{l^{\prime}}^{2}-q_{l^{\prime\prime}}^{2}\right)  \;.
\label{ql' / ql''}%
\end{equation}
The above relation (\ref{ql' / ql''}) implies that the ratio $q_{l^{\prime}%
}/q_{l^{\prime\prime}}$ monotonically increases with time, provided that
$q_{l^{\prime}}>q_{l^{\prime\prime}}$ (recall that $\gamma>0$). This behavior
gives rise to disentanglement. Consider the case where initially, at time
$t=0$, $q_{l_{0}}=\max\left\{  q_{l}\right\}  $ for a unique positive integer
$l_{0}\in\left\{  1,2,\cdots,\min\left(  N_{1},N_{2}\right)  \right\}  $. For
this case, $\left\vert \psi\right\rangle $\ evolves into the product state
$\left\vert l_{0},l_{0}\right\rangle $ in the long time limit, i.e.
$q_{l}\rightarrow\delta_{l,l_{0}}$\ in the limit $t\rightarrow\infty$ (see
Fig. \ref{Figdqdt}). Note, however, that in the long time limit the state can
be strongly affected by noise when initially the set $\left\{  q_{l}\right\}
$ doesn't have a unique member significantly larger than all others.

\textbf{Discussion} - As was already mentioned above, several types of
nonlinear extensions of quantum mechanics have been proposed and explored
\cite{Bassi_471,Bennett_170502,Kowalski_1,Fernengel_385701,Kowalski_167955}.
However, it was found that for some cases, the proposed nonlinear extension
gives rise to the violation of the causality principle by enabling
superluminal signaling
\cite{Bassi_055027,Jordan_022101,Polchinski_397,Helou_012021}. More recently,
it was shown that when a condition called 'convex quasilinearity' is satisfied
by a given nonlinear master equation, the violation of the causality principle
becomes impossible \cite{Rembielinski_012027,Rembielinski_420}. Some of the
proposed nonlinear extensions are inconsistent with the principle of
separability \cite{Hejlesen_thesis,Jordan_022101,Jordan_012010}. Moreover, any
proposed extension must be ruled out if it alters predictions of standard
quantum mechanics that have been experimentally confirmed.

The modified Schr\"{o}dinger equation given by Eq. (\ref{NLSE}) has an
important advantage compared to other proposals: the added nonlinear term
$-\gamma\left(  \mathcal{Q}-\left\langle \mathcal{Q}\right\rangle \right)  $
has no effect on product states. This implies that in the absence of
entanglement, the added term does not vary any prediction of standard quantum
mechanics. Moreover, possible conflicts with both principles of causality and
separability can be avoided by postulating that $\gamma\simeq\hbar
^{-1}\left\langle \psi\right\vert V^{\dag}V^{{}}\left\vert \psi\right\rangle
^{1/2}$, where $V$ is the coupling term in the Hamiltonian giving rise to the
interaction between subsystems [$\gamma$ is the disentanglement rate in Eq.
(\ref{NLSE})]. This postulate implies that the added nonlinear term is active
only when subsystems interact, and that time evolution is governed by the
standard Schr\"{o}dinger equation when subsystems are remote (i.e. decoupled).
Note that for the examples shown in Fig. \ref{Fig2SpinDE}, the calculations
are performed for the case $\gamma=\omega_{\mathrm{d}}$. This demonstrates
that a disentanglement rate $\gamma$ having the order of $\hbar^{-1}%
\left\langle \psi\right\vert V^{\dag}V^{{}}\left\vert \psi\right\rangle
^{1/2}$ is sufficiently large to allow full suppression of entanglement.

\textbf{Summary} - Further theoretical study is needed to check whether
quantum mechanics can be self-consistently reformulated based on the proposed
modified Schr\"{o}dinger equation (\ref{NLSE}). We find that conflict with
some well-established physical principles, as well as many experimental
observations, can be avoided by postulating that $\gamma\simeq\hbar
^{-1}\left\langle \psi\right\vert V^{\dag}V^{{}}\left\vert \psi\right\rangle
^{1/2}$.

The expression given by Eq. (\ref{Q operator}) for the operator $\mathcal{Q}$
is applicable for the bipartite case, for which the entire system is divided
into two subsystems. The multipartite case, however, for which the entire
system is divided into more than two subsystems, requires a generalization of
Eq. (\ref{Q operator}). Such generalization is discussed in Ref.
\cite{Buks_2306_05853}. The generalization of the above discussed postulate
(regarding the disentanglement rate $\gamma$) for the multipartite case states
that disentanglement between two given subsystems is active only during the
time when they interact.

Further insight can be gained from experimental study of entanglement in the
region where environmental decoherence is negligible \cite{Buks_014421}. Upper
bounds imposed upon the disentanglement rate $\gamma$ in Eq. (\ref{NLSE}) can
be derived from lifetime measurements of entangled states. Experimental
observations of deviation from the Born rule may provide supporting evidence
for nonlinearity (see Fig. \ref{FigN}).

\textbf{Acknowledgments} - We thank Jakub Rembielinski, Pawel\ Caban, Joakim
Bergli and Klaus Molmer for useful discussions. This work was supported by the
Israeli science foundation, the Israeli ministry of science, and by the
Technion security research foundation.

\appendix

\section{The Schmidt decomposition}

The system's normalized pure state vector $\left\vert \psi\right\rangle $ is
given by $\left\vert \psi\right\rangle =\mathcal{K}_{1}C\otimes\mathcal{K}%
_{2}^{\mathrm{T}}$ [see Eq. (\ref{psi C}) in the main text]. Consider the
unitary transformations (the letter $k$ is used to label the states of the
original basis, whereas the transformed states are labeled by the letter $l$)%
\begin{align}
\mathcal{K}_{1}^{\mathrm{T}}  &  =u_{1}\mathcal{L}_{1}^{\mathrm{T}}%
=u_{1}\left(  \left\vert l_{1}\right\rangle _{1},\left\vert l_{2}\right\rangle
_{1},\cdots,\left\vert l_{N_{1}}\right\rangle _{1}\right)  ^{\mathrm{T}}\;,\\
\mathcal{K}_{2}^{\mathrm{T}}  &  =u_{2}\mathcal{L}_{2}^{\mathrm{T}}%
=u_{2}\left(  \left\vert l_{1}\right\rangle _{2},\left\vert l_{2}\right\rangle
_{2},\cdots,\left\vert l_{N_{2}}\right\rangle _{2}\right)  ^{\mathrm{T}}\;,
\end{align}
where $u_{1}$ ($u_{2}$) is a $N_{1}\times N_{1}$ ($N_{2}\times N_{2}$) unitary
matrix (i.e. $u_{1}^{\dag}u_{1}=1$ and $u_{2}^{\dag}u_{2}=1$). The state
vector $\left\vert \psi\right\rangle $ in the transformed basis is expressed
as%
\begin{align}
\left\vert \psi\right\rangle  &  =\mathcal{L}_{1}\hat{C}\otimes\mathcal{L}%
_{2}^{\mathrm{T}}\nonumber\\
&  =%
{\displaystyle\sum\limits_{l_{1},l_{2}}}
\hat{C}_{l_{1},l_{2}}\left\vert l_{1}\right\rangle _{1}\otimes\left\vert
l_{2}\right\rangle _{2}\;,\nonumber\\
&  \label{|xi> L}%
\end{align}
where the transformed matrix $\hat{C}$ is given by%
\begin{equation}
\hat{C}=u_{1}^{\mathrm{T}}Cu_{2}\;,
\end{equation}
and the corresponding density operator $\rho=\left\vert \psi\right\rangle
\left\langle \psi\right\vert $ is expressed as%
\begin{equation}
\rho=%
{\displaystyle\sum\limits_{l_{1}^{\prime},l_{2}^{\prime},l_{1}^{\prime\prime
},l_{2}^{\prime\prime}}}
\hat{C}_{l_{1}^{\prime},l_{2}^{\prime}}^{{}}\hat{C}_{l_{1}^{\prime\prime
},l_{2}^{\prime\prime}}^{\ast}\left\vert l_{1}^{\prime},l_{2}^{\prime
}\right\rangle \left\langle l_{1}^{\prime\prime},l_{2}^{\prime\prime
}\right\vert \;. \label{rho C hat}%
\end{equation}
The following holds%
\begin{align}
\operatorname{Tr}\rho &  =%
{\displaystyle\sum\limits_{l_{1},l_{2}}}
\left\vert \hat{C}_{l_{1},l_{2}}\right\vert ^{2}\nonumber\\
&  =\operatorname{Tr}S_{1}=\operatorname{Tr}S_{2}=\operatorname{Tr}\left(
CC^{\dag}\right)  =\operatorname{Tr}\left(  C^{\dag}C\right)  \;,\nonumber\\
&  \label{Tr rho SD}%
\end{align}
where the $N_{1}\times N_{1}$ ($N_{2}\times N_{2}$) matrix $S_{1}$ ($S_{2}$)
is given by (recall that $u_{1}^{\dag}u_{1}=1$ and $u_{2}^{\dag}u_{2}=1$)%
\begin{align}
S_{1}  &  =\hat{C}\hat{C}^{\dag}=u_{1}^{\mathrm{T}}Cu_{2}u_{2}^{\dag}C^{\dag
}u_{1}^{\mathrm{T\dag}}=u_{1}^{\mathrm{T}}CC^{\dag}u_{1}^{\mathrm{T\dag}}\;,\\
S_{2}  &  =\hat{C}^{\dag}\hat{C}=u_{2}^{\dag}C^{\dag}u_{1}^{\mathrm{T\dag}%
}u_{1}^{\mathrm{T}}Cu_{2}=u_{2}^{\dag}C^{\dag}Cu_{2}\;,
\end{align}
hence $\operatorname{Tr}\rho=1$ provided that $\left\vert \psi\right\rangle $
is normalized. The matrix $S_{1}$ ($S_{2}$) is Hermitian and positive
definite, hence the unitary matrix $u_{1}$ ($u_{2}$) can be chosen to
diagonalize $S_{1}$ ($S_{2}$), and the eigenvalues, which are denoted by
$q_{l}$, are non-negative. For this transformation, which is called the
Schmidt decomposition, the transformed matrix $\hat{C}$ has a diagonal form%
\begin{equation}
\hat{C}_{l_{1},l_{2}}=q_{l_{1}}\delta_{l_{1},l_{2}}\;. \label{C Schmidt}%
\end{equation}

The purity $P_{1}$ ($P_{2}$) is defined by $P_{1}=\operatorname{Tr}\rho
_{1}^{2}$ ($P_{2}=\operatorname{Tr}\rho_{2}^{2}$), where $\rho_{1}%
=\operatorname{Tr}_{2}\rho$\ ($\rho_{2}=\operatorname{Tr}_{1}\rho$) is the
reduced density operator of the first (second) subsystem. With the help of the
Schmidt decomposition (\ref{C Schmidt}), one finds that $P_{1}=P_{2}\equiv P$,
where%
\begin{align}
P  &  =%
{\displaystyle\sum\limits_{l}}
q_{l}^{4}\nonumber\\
&  =\operatorname{Tr}S_{1}^{2}=\operatorname{Tr}\left(  CC^{\dag}\right)
^{2}=\operatorname{Tr}S_{2}^{2}=\operatorname{Tr}\left(  C^{\dag}C\right)
^{2}\;.\nonumber\\
&
\end{align}
Note that $P=1\ $for a product state, and $P$ obtains its minimum value of
$1/\min\left(  N_{1},N_{2}\right)  $ for a maximally entangled state. The
purity $P$ is independent on the local transformations $u_{1}$ and $u_{2}$,
hence it is a constant when the subsystems are decoupled (i.e. when the
interaction between the subsystems vanishes). Using the relations%
\begin{equation}
\operatorname{Tr}\left(  C^{\dag}C\right)  =%
{\displaystyle\sum\limits_{k_{1}^{\prime}=1}^{N_{1}}}
{\displaystyle\sum\limits_{k_{2}^{\prime}=1}^{N_{2}}}
C_{_{k_{1}^{\prime},k_{2}^{\prime}}}^{\ast}C_{_{k_{1}^{\prime},k_{2}^{\prime}%
}}\;,
\end{equation}
and%
\begin{align}
\operatorname{Tr}\left(  C^{\dag}C\right)  ^{2}  &  =%
{\displaystyle\sum\limits_{k_{1}^{\prime},k_{1}^{\prime\prime}=1}^{N_{1}}}
{\displaystyle\sum\limits_{k_{2}^{\prime},k_{2}^{\prime\prime}=1}^{N_{2}}}
C_{k_{1}^{\prime},k_{2}^{\prime}}^{\ast}C_{k_{1}^{\prime},k_{2}^{\prime\prime
}}C_{k_{1}^{\prime\prime},k_{2}^{\prime\prime}}^{\ast}C_{k_{1}^{\prime\prime
},k_{2}^{\prime}}\;,\nonumber\\
&
\end{align}
one finds that the level of entanglement $1-P$ is given by%
\begin{align}
1-P  &  =\left(  \operatorname{Tr}\left(  C^{\dag}C\right)  \right)
^{2}-\operatorname{Tr}\left(  C^{\dag}C\right)  ^{2}\nonumber\\
&  =\frac{1}{2}%
{\displaystyle\sum\limits_{k_{1}^{\prime},k_{1}^{\prime\prime}=1}^{N_{1}}}
{\displaystyle\sum\limits_{k_{2}^{\prime},k_{2}^{\prime\prime}=1}^{N_{2}}}
\left\vert \phi_{k_{1}^{\prime},k_{1}^{\prime\prime},k_{2}^{\prime}%
,k_{2}^{\prime\prime}}\right\vert ^{2}\;,\nonumber\\
&  \label{1-P V1}%
\end{align}
where%
\begin{equation}
\phi_{k_{1}^{\prime},k_{1}^{\prime\prime},k_{2}^{\prime},k_{2}^{\prime\prime}%
}=C_{_{k_{1}^{\prime},k_{2}^{\prime}}}C_{_{k_{1}^{\prime\prime},k_{2}%
^{\prime\prime}}}-C_{k_{1}^{\prime},k_{2}^{\prime\prime}}C_{k_{1}%
^{\prime\prime},k_{2}^{\prime}}\;. \label{psi k prime}%
\end{equation}
Note that the term $\phi_{k_{1}^{\prime},k_{1}^{\prime\prime},k_{2}^{\prime
},k_{2}^{\prime\prime}}$ vanishes unless $k_{1}^{\prime}\neq k_{1}%
^{\prime\prime}$ and $k_{2}^{\prime}\neq k_{2}^{\prime\prime}$, and the
following holds $\phi_{k_{1}^{\prime},k_{1}^{\prime\prime},k_{2}^{\prime
},k_{2}^{\prime\prime}}=\phi_{k_{1}^{\prime\prime},k_{1}^{\prime}%
,k_{2}^{\prime\prime},k_{2}^{\prime}}$, thus Eq. (\ref{1-P V1}) can be
rewritten as%
\begin{equation}
1-P=2%
{\displaystyle\sum\limits_{k_{1}^{\prime}<k_{1}^{\prime\prime}}}
{\displaystyle\sum\limits_{k_{2}^{\prime}<k_{2}^{\prime\prime}}}
\left\vert \phi_{k_{1}^{\prime},k_{1}^{\prime\prime},k_{2}^{\prime}%
,k_{2}^{\prime\prime}}\right\vert ^{2}\;. \label{1-P V2}%
\end{equation}
Note that for any product state $\phi_{k_{1}^{\prime},k_{1}^{\prime\prime
},k_{2}^{\prime},k_{2}^{\prime\prime}}=0$ [see Eq. (\ref{psi k prime})]. The
above result (\ref{1-P V2}) implies that $P=1-\left\langle \mathcal{Q}%
\right\rangle $, where the operator $\mathcal{Q}$ is given by Eq.
(\ref{Q operator}) in the main text.

\bibliographystyle{ieeepes}
\bibliography{acompat,Eyal_Bib}

\end{document}